\documentclass[preprint,12pt]{elsarticle}

\usepackage{hyperref}

\journal{XXX}

\usepackage{graphicx, color, transparent}
\usepackage{euscript}
\usepackage{amsmath, amsfonts}
\usepackage{breqn}
\usepackage{setspace}
\usepackage{bm}
\usepackage[dvipsnames]{xcolor}

\usepackage[normalem]{ulem}

\begin{document}

\begin{frontmatter}

\title{On local kirigami mechanics II: Stretchable creased solutions}

\author[1]{Souhayl Sadik\corref{cor}}
\ead{sosa@mpe.au.dk}

\author[2]{Martin G. Walker}
\ead{m.g.walker@surrey.ac.uk}

\author[3]{Marcelo A. Dias\corref{cor}}
\cortext[cor]{Corresponding authors}
\ead{marcelo.dias@ed.ac.uk}

\address[1]{Department of Mechanical and Production Engineering, Aarhus University, 8000~Aarhus~C, Denmark}
\address[2]{Department of Civil and Environmental Engineering, University of Surrey, Surrey~GU2~7XH, UK}
\address[3]{Institute for Infrastructure \& Environment, School of Engineering, The University of Edinburgh, Edinburgh~EH9~3FG Scotland, UK}

\begin{abstract}
Following on Part I of this work series on local kirigami mechanics, we present a study of a discretely creased mechanism as a model to investigate the mechanics of the basic geometric building block of kirigami\textemdash the e-cone. We consider an annular disk with a single radial slit discritised by a series of radial creases connecting kinematically flat rigid panels. The creases allow both relative rotation and separation between panels, capturing both bending and stretching deformations. Admissible equilibrium configurations are obtained by penalising these deformations using elastic springs with stiffnesses derived from compatible continuum plate deformations. This provides a tool to study both inextensible and extensible e-cone configurations due to opening of the slit and rotation of its lips. This creased model hence offers the possibility to study the e-cone away from its isometric limit, i.e., for plates with finite thickness, and explore the full range of post-buckling (far-from-threshold) behaviour as well as initial buckling (near-threshold) instability. Our local approach provides a fundamental understanding of kirigami phenomenology, underpinned by a proper theoretical approach to geometry and mechanics.
\end{abstract}


\end{frontmatter}


\newcommand{\sosa}[1]{\textcolor{blue}{\emph{[SoSa: ---#1]}}}
\newcommand{\mad}[1]{\textcolor{red}{\emph{[MAD: ---#1]}}}
\newcommand{\mgw}[1]{\textcolor{SeaGreen}{\emph{[MGW: ---#1]}}}
\definecolor{pk1}{rgb}{.88, 0.36, 0.50}
\section{Introduction}
\label{sec:1}

Present-day challenges in engineering continue to benefit from ancient ideas. This is the case for morphing and deployable structures~\citep{Saito2011,Lamoureux2015}, highly stretchable graphene sheets and electronic devices~\citep{Qi2014,Blees2015,Zhang2015,Song2015,Han2017}, nanocomposites~\citep{Shyu2015,Xu2016}, MEMS~\cite{Rogers2016,Baldwin2017}, and tuneable tribological properties~\citep{rafsanjani2018kirigami}, just to name a few examples.
What these engineering advances all have in common is their inspiration from the Japanese art forms of paper folding and paper cutting, respectively, origami and kirigami. These have become useful exercises as they help to unlock novel strategies to design a material's kinematic degrees of freedom as well as its effective mechanical response---all achievable through changes in geometry and topology via the insertion of folds and cuts~\cite{Castle2014,Sussman2015}. In particular, kirigami has recently been gaining attention because of its fascinating complementary properties to origami. While introducing cuts or slits purposely into a medium may seem counter-intuitive, given their susceptibility to failure through fracture processes, kirigami offers further fundamental motifs of deformation in thin elastic sheets and unlocks new functionality. Cuts under tension induce a local build-up of compressive stresses~\citep{Dias2017b,mahmood2018cracks}, which in turn can be used to manipulate a thin sheet's topographic profile and program its mechanical behaviour~\citep{Scarpa2013,Cai2016,dias2018,moshe2019nonlinear,moshe2019kirigami}. Other fundamental advantages of kirigami are its ability to regulate stretchability~\citep{isobe2019continuity, rafsanjani2019propagation} and enable precise manipulation of shape-changing structures at a variety of scales~\citep{Dias2017b,kaspersen2019lifting,alderete2021programmable,zhang2021kirigami}. As we continue to seek potential applications inspired by kirigami, progress on the fundamental understanding of such structures is still needed. Efforts to derive analytical models breaking down the multi-scale and non-linear phenomenology of kirigami, underpinned by a proper theoretical approach to geometry and mechanics, certainly can not be ignored if we aim at better designs and reduced computational costs. This is indeed the target of this work series; to propose a detailed theoretical account of local mechanical behaviour of kirigami-inspired materials and structures.

\begin{figure}
\centering
\def\svgwidth{.75\textwidth}
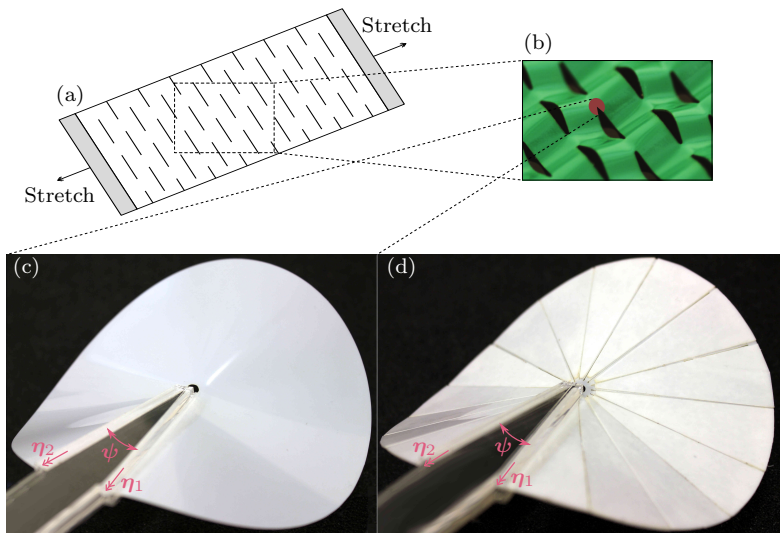
\caption{Kirigami consists of making cuts in a flat sheet (a). The placement of such cuts endows the sheet with a higher capacity for stretching thanks to local out-of-plane buckling (b). Such buckling reduces to e-cone motifs locally forming around the tip of every cut, hence being the basic building bloc of kirigami. We show e-cones as a smoothly deformed surface (c) and a discretely deformed creased disk (d).}
\label{fig:Kiri2Econe}       
\end{figure}

Following Part I of this work series~\cite{sadik2021kiriI}, we continue to focus on the fundamental local mechanics of kirigami through its unit motif. As shown in Fig.~\ref{fig:Kiri2Econe}-(a-c), this motif is a disclination-like defect in a medium, where global geometry is dictated by localised sources of negative Gaussian curvature. Known as an e-cone~\citep{muller2008conical,Guven2013,efrati2015confined,seffen2016fundamental}, this feature constitutes the basic geometric building bloc of kirigami; Fig.~\ref{fig:Kiri2Econe}-(c). In part I, we presented an analytical post-buckling study of a nonlinear continuum plate model for the e-cone in its isometric limit. As the thickness of the sheet approaches zero, the stretching energy contribution vanishes and the plate undergoes an inextensible deformation governed solely by the bending energy.
In the present work, the main objective is to propose a model for local kirigami mechanics that moves beyond the isometric zero thickness limit. Following an approach proposed by Walker in the study of creased disks~\cite{walker2020mechanics}, we swap the nonlinear continuum plate model for a kinematic mechanism\textemdash see Fig.~\ref{fig:Kiri2Econe}-(d)\textemdash which replaces smooth deformation by foldable and separable radial sharp creases connecting rigid panels. In this model, bending is accounted for by axial rotational springs along foldable hinges at the creases, and stretching by circumferential spring joints allowing separation of the panels at the creases. The mechanics is naturally derived from the continuum such that the spring constants are found by equating the elastic energies of the hinges and joints to the bending and stretching energies of the panels, respectively.
This is a powerful method, as it fully captures the entire range of the far-from-threshold post-buckling behaviour previously presented in~\cite{sadik2021kiriI}, while opening up the possibility to study the near-threshold behaviour of instability. Moreover, the appearance of creases in e-cones is more than a model, as it suggests an amalgamation of kirigami and origami. Indeed, the introduction of additional creases to accommodate non-rigid panel bending in origami structures is well established~\cite{Tachi2013,Evans2015,Liu2017,andrade2019foldable,yu2021cutting}. Such reduced-order bending models have a long history in engineering having been applied to phenomena ranging from the buckling and collapse of thin-walled structures~\cite{Zhao2003,Hiriyur2005} to the analysis of reinforced concrete slabs~\cite{Kennedy2004}. Our approach retains the computational efficiency of reduced-order models while maintaining the capacity to describe the continuum.

The manuscript is organised as follows.
In \S\ref{sec:2}, we present the creased model of the disk. We set out the underlying mechanism, lay out its kinematic description, and derive its mechanics from the F{\"o}ppl-von K{\'a}rm{\'a}n plate equations.
In \S\ref{sec:3}, we show how the creased model rapidly approaches the analytical solution in the isometric limit. presented in Part I, as the number of creases is increased.
Looking beyond the isometric limit, in \S\ref{sec:4} we include the stretching contribution for finite-thickness plates. We study the effect of stretching on the initial buckling instability (near-threshold behaviour) of the e-cone and the role it plays in the non-linear post-buckling paths and further loss of stability (far-from-threshold behaviour). We show how the onset of buckling is delayed for thicker plates, how the strain energy equipartition is affected, and how the Gaussian charge at the apex, imposed as a result of the excess angle, is not entirely realised into bending of the e-cone.
In \S\ref{sec:4}, we close with a few concluding remarks.

\section{Discrete creased e-cone model}
\label{sec:2}

We consider a thin, initially planar annular disk of thickness $h\,$, inner radius $R_i\,$, and outer radius $R_o$ with a radial slit extending from the centre to the boundary. We concern ourselves with the study of its deformation following the opening of the slit with an excess angle $\psi$ and the rotation of its lips with angles $\eta_1$ and $\eta_2\,$. Experimentally deformed smooth and creased e-cones can be seen in Fig.~\ref{fig:Kiri2Econe}-(c)~\&~(d), respectively. First, we present the kinematics of the creased disk mechanism, then we derive the mechanics, thereby providing a complete mechanical description of the creased disk.

\subsection{Kinematics}

\begin{figure}[t]
\centering
\def\svgwidth{\textwidth}
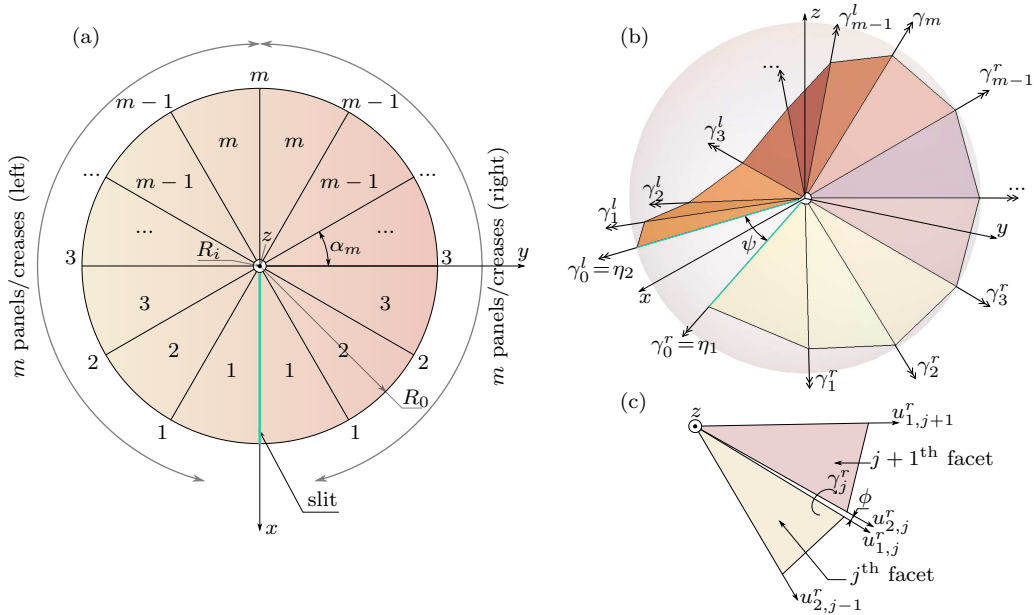
\caption{Schematic diagrams for a creased annulus with inner radius $R_i$ and outer radius $R_o\,$. (a) shows the creases layout, dividing each half-annulus into $m$ facets of sector angle ${\alpha_m=\pi/m}\,$. The creases are numbered symmetrically, starting with the slit as $0$ to $m$ for the antipodal crease. (b) shows the annulus in the deformed configuration and indicates the rotation vectors between facets, $\gamma_i\,$, indicated as positive by a radially outward pointing vectors according to the right hand rule. Superscripts $l$ and $r$ indicate rotations on the left or right side of the annulus respectively. The two sides of the slit are opened to an angle of $\psi$ and the imposed rotation angles at the slit lips labeled by $\eta_1 = \gamma_0^r$ and $\eta_2 = \gamma_0^l\,$. (c) shows the labelling convention for the edge vectors of each facet.  Edge vectors, $u_{1,n}$ and $u_{2,n}\,$, are numbered in the clockwise direction on the right side and anti-clockwise on the left side, where $n$ is the crease number\textemdash see (a). The superscript indicates which side (left - l, right - r) the facet is located on.}
\label{fig:crease_mod}       
\end{figure}

Consider a discretised disk made of $2m$ identical panels connected by $2m-1$ radial creases uniformly spaced by an angle of $\alpha_m=\pi/m\,$, where $m$ is the discretisation size. We label the two half-disks on either side of the slit as right and left such that the right half corresponds to the lip rotated by $\eta_1$ and the left half to the one rotated by $\eta_2\,$. Starting from the slit and ending on the antipodal crease, panels and creases are numbered in increasing order on either half-disk\textemdash see Fig.~\ref{fig:crease_mod}-(a).
Each radial crease acts both as a hinge (to model bending by allowing relative rotation of the panels around its axis\textemdash Fig.~\ref{fig:crease_mod}-(b)) and as a joint (to model stretching by allowing relative angular rotation of the panels about the $z$ axis\textemdash Fig.~\ref{fig:crease_mod}-(c)).

Let $\{x,y,z\}$ be an orthonormal Cartesian coordinate system with its origin attached to the centre of the disk such that, in the disk's planar undeformed state, the $x$-axis is along the slit, the $z$-axis is normal to the disk, and the $y$-axis points towards the right half disk\textemdash see Fig.~\ref{fig:crease_mod}-(a). As the slit of the disk opens, its lips are constrained to remain on the $xy$-plane such that $x$-axis bisects the excess angle they form.

The lips' rotation angles are denoted by $\gamma^r_0=\eta_1$ and $\gamma^l_0=\eta_2\,$. For $j \in \{1,...,m\}\,$, we denote the relative hinge rotation angle of the adjacent panels joined by the $j^{\textrm{th}}$ right (left, respectively) crease by $\gamma^r_j$ ($\gamma^l_j\,$)\textemdash defined as positive by the right-hand rule with respect to the radially outward unit vector along the hinge line. At the $m^{\textrm{th}}$ crease, we have $\gamma_m = \gamma^r_m = \gamma^l_m\,$. See Fig.~\ref{fig:crease_mod}-(b).
We assume the relative separation angle between adjacent panels, at the joint connecting them, to be uniform and denote it by $\phi\,$. This is effectively a relative circumferential rotation of the adjacent panels around the $z$-axis such that $\phi>0$ for a gap in case of tension, and $\phi<0$ for an overlap in case of compression.

For $j \in \{1,...,m\}\,$, we denote the radially outward unit vectors along the edges of the panels on either side of the joint on the $j^{\textrm{th}}$ right (left, respectively) crease by $\boldsymbol u^r_{1,j}$ and $\boldsymbol u^r_{2,j}$ ($\boldsymbol u^l_{1,j}$ and $\boldsymbol u^l_{2,j}\,$, respectively), according to the crease numbering\textemdash see Fig.~\ref{fig:crease_mod}-(c). On the right (left) lip of the slit, we denote the radially outward unit vector along it by $\boldsymbol u^r_{1,0} = \boldsymbol u^r_{2,0}$  ($\boldsymbol u^l_{1,0}=\boldsymbol u^l_{2,0}\,$).
We let $\boldsymbol n^r_0=\boldsymbol n^l_0$ be the unit normal to the disk in its initial planar state along the $z$-axis and we denote by $\boldsymbol n^r_j$ ($\boldsymbol n^l_j\,$, respectively) for $j \in \{1,...,m\}\,$, the normal to the $j^{\textrm{th}}$ right (left) panel such that the ordered triad $\{\boldsymbol u^r_{2,j-1}, \boldsymbol u^r_{1,j}, \boldsymbol n^r_j\}$ ($\{\boldsymbol u^l_{1,j}, \boldsymbol u^l_{2,j-1}, \boldsymbol n^l_j\}\,$, respectively) is positively oriented following the right-hand rule.
In the disk's deformed configuration, the creases and normal unit vectors may be expressed iteratively for all $j \in \{1,...,m\}$ as follows
\begin{subequations}
\allowdisplaybreaks
\label{kin_rec}
\begin{align}
&\left\{{\setstretch{1.3}\arraycolsep=1.75pt
\begin{array}{ll}
\boldsymbol u^r_{1,0}&= \boldsymbol u^r_{2,0} =\left(\cos(\psi/2) ~ \sin(\psi/2) ~ 0\right)^{\mathsf T}\!,\\
\boldsymbol n^r_0 &=\left(0 \; 0 \; 1 \right)^{\mathsf T}\!,~
\gamma^r_0 = \eta_1\,,\\
\boldsymbol u^r_{1,j} &= \sin\left({\pi}/{m}\right)\left[\sin(\gamma^r_{j-1})\boldsymbol n^r_{j-1}+\cos(\gamma^r_{j-1}) \boldsymbol n^r_{j-1}\times \boldsymbol u^r_{2,j-1}\right]\\
& \quad+\cos\left({\pi}/{m}\right) \boldsymbol u^r_{2,j-1},\\
\boldsymbol u^r_{2,j} &= R_{\hat{\boldsymbol e}_z}(\phi).\boldsymbol u^r_{1,j},\\
\boldsymbol n^r_j &= \boldsymbol u^r_{2,j-1} \times \boldsymbol u^r_{1,j}/\sin\left({\pi}/{m}\right)\,,
\end{array}}\right.
\\[5pt]
&\left\{{\setstretch{1.3}\arraycolsep=1.75pt
\begin{array}{ll}
\boldsymbol u^l_{1,0} &=\boldsymbol u^l_{2,0}=\left(\cos(\psi/2) ~ -\!\sin(\psi/2) ~ 0 \right)^{\mathsf T}\!,\\
\boldsymbol n^l_0 &=\left( 0 ~ 0 ~ 1 \right)^{\mathsf T}\!,~
\gamma^l_0 = \eta_2\,,\\
\boldsymbol u^l_{1,j} &= \sin\left({\pi}/{m}\right)\big[\sin(\gamma^l_{j-1})\boldsymbol n^l_{j-1}-\cos(\gamma^l_{j-1})\boldsymbol n^l_{j-1}\times \boldsymbol u^l_{2,j-1}\big]\\
& \quad+\cos\left({\pi}/{m}\right) \boldsymbol u^l_{2,j-1}\,,\\
\boldsymbol u^l_{2,j} &= R_{\hat{\boldsymbol e}_z}(\phi).\boldsymbol u^l_{1,j},\\
\boldsymbol n^l_j &= -\boldsymbol u^l_{2,j-1} \times \boldsymbol u^l_{1,j}/\sin\left({\pi}/{m}\right)\,,
\end{array}}\right.
\end{align}
\end{subequations}
where $R_{\hat{\boldsymbol e}_z}(\phi)$ denotes the rotation matrix around the $z$-axis unit vector $\hat{\boldsymbol e}_z$ by the angle $\phi\,$.
To ensure continuity of the kinematic description between the left and right halves of the disk at the crease anti-podal to the slit, i.e., the $m^{\textrm{th}}$ crease, we require
\begin{equation}
\label{kin_con}
\boldsymbol u^r_{1,m} = \boldsymbol u^l_{2,m}\,\quad\textrm{and}\quad
\boldsymbol u^r_{2,m} = \boldsymbol u^l_{1,m}\,.
\end{equation}
Following the kinematic description laid out above, given the lip rotations $\gamma^r_0=\eta_1$ and $\gamma^l_0=\eta_2\,$, infinitely many configurations of the e-cone may be realised by the recursive scheme given in Eq.~\eqref{kin_rec} for any arbitrary choice of hinge rotations
$\{\gamma^r_1,...,\gamma^r_{m-1},\gamma^l_1,...,\gamma^l_{m-1}\}$ and a joint angular displacement $\phi\,$. For such a set of angles to constitute a kinematically compatible deformation, it must  satisfy the continuity conditions given by Eq.~\eqref{kin_con}. Note that the hinge rotation $\gamma_m\,$, at the $m^{\textrm{th}}$ crease, need not be further specified: it is the angle between $\boldsymbol n^r_m$ and $\boldsymbol n^l_m\,$, which are recursively obtained by Eq.~\eqref{kin_rec} from the joint opening and the hinge rotation angles excluding $\gamma^m\,$.

In the next section, we present a mechanics description of the creased disk model by energetically penalising the deformation at the creases and derive the associated total strain energy.

\subsection{Mechanics}

In the discretely creased e-cone model introduced in the preceding section, the creases have two crucial features: as hinges, they allow relative rotation of any two adjacent panels (around the radial hinge lines) to model bending; and as joints, they allow relative angular displacement of any two adjacent panels (around the $z$-axis) to model stretching.

In order to identify the admissible equilibrium configurations of the disk, the kinematic description needs to be complemented by the mechanics of the model. At the creases, we attach rotational springs, with bending stiffness $k_b\,$, to the hinges (as a discretised model for bending); and circumferential rotational springs, with stretching stiffness $k_s\,$, to the joints (as a discretised model for stretching). We then write the elastic energy $\mathcal W$ of the disk as the sum of a bending contribution $\mathcal W_b$ with respect to the hinge rotations $\{\gamma^r_1,...,\gamma^r_{m-1},\gamma^l_1,...,\gamma^l_{m-1},\gamma_m\}$ penalised by $k_b$ and a stretching contribution $\mathcal W_s$ with respect to the uniform joint relative angular displacement $\phi$ penalised by $k_s$; it reads

\begin{equation}
\label{tot_E}
\mathcal W = \underbrace{\frac{1}{2} k_b  \left\{\sum_{j=1}^{m-1} \left[\left(\gamma^r_j\right)^2+\left(\gamma^l_j\right)^2\right] + \left(\gamma_m\right)^2\right\}}_{\mathcal W_b}\quad+\underbrace{\frac{2m-1}{2} k_s \phi^2}_{\mathcal W_s}\,.
\end{equation}

In order to suitably approximate the continuum behaviour of the e-cone, we relate the hinge and joint stiffnesses ($k_b$ and $k_s$) to the mechanics of the panels they represent. We consider a wedge-shaped elastic plate with the same elastic properties as the disk and the same geometry as that of the rigid panels forming the creased disk. This elastic plate is subject to a rotation on either edge equal to half the hinge rotation, i.e., $\gamma/2\,$, and to a stretch changing its subtended angle from $\alpha_m$ to $\alpha_m+\phi$\textemdash these boundary conditions are shown in Fig.~\ref{fig:elast_pan}. The crease stiffnesses are then found by equating the elastic energies of the hinge ($k_b \gamma^2/2$) and joint ($k_s \phi^2/2$) to the bending and stretching energies of the plate, respectively. These energies are computed by solving for continuum degrees-of-freedom of the elastic plate. We let $(r,\theta)$ be a polar coordinate system for the panel such that its origin is at the apex. The wedge has a fixed lateral side at $\theta=0$ and a free side at $\theta=\alpha_m\,$. We solve for the deformed equilibrium solution of the plate using the F{\"o}ppl-von K{\'a}rm{\'a}n equations: 

\begin{figure}
\centering
\def\svgwidth{.35\textwidth}
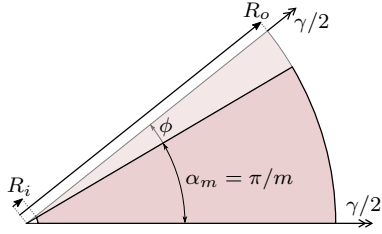
\caption{Elastic panel deformation: a rotation on either edge equal to half the hinge rotation, i.e., $\gamma/2\,$, and a stretch by changing its subtended angle from $\alpha_m$ to $\alpha_m+\phi$}
\label{fig:elast_pan}       
\end{figure}

\begin{subequations}
\label{eq:FvK}
\begin{align}
    \frac{1}{r}\frac{\partial }{\partial r}(r \sigma_{rr})-\frac{\sigma_{\theta\theta}}{r}+\frac{1}{2}\frac{\partial \sigma_{r\theta}}{\partial\theta}=0\,, \label{eq:FvK1}\\
    \frac{1}{r}\frac{\partial }{\partial r}(r \sigma_{r\theta})+\frac{\sigma_{r\theta}}{r}+\frac{1}{r}\frac{\partial \sigma_{\theta\theta}}{\partial\theta}=0\,, \label{eq:FvK2}\\
    D\nabla^4 w - h\left[\sigma_{rr} \frac{\partial^2 w}{\partial r^2}+2\sigma_{r\theta}\frac{\partial }{\partial r}\left(\frac{1}{r}\frac{\partial w}{\partial \theta}\right)\!+\sigma_{\theta\theta}\left(\frac{1}{r}\frac{\partial w}{\partial r} +\frac{1}{r^2}\frac{\partial^2 w}{\partial \theta^2}\right)\right] =0\,, \label{eq:FvK3}
\end{align}
\end{subequations}
where the biharmonic operator in polar coordinates reads as follows
\begin{equation}
\begin{split}
\nabla^4 w=&\frac{1}{r}\frac{\partial}{\partial r}\left\{r\frac{\partial}{\partial r}\left[\frac{1}{r} \frac{\partial}{\partial r}\left(r \frac{\partial w}{\partial r}\right)\right]\right\}+\frac{2}{r^2}\frac{\partial^4 w}{\partial \theta^2 \partial r^2}+\frac{1}{r^4}\frac{\partial^4 w}{\partial \theta^4}\\&-\frac{2}{r^3}\frac{\partial^3 w}{\partial \theta^2 \partial r}-\frac{4}{r^4}\frac{\partial^2 w}{\partial \theta^2}\,,
\end{split}
\end{equation}
and the components of the strain tensor $\bm{\epsilon}$ in cylindrical coordinates are given by
\begin{equation}
\begin{split}
    \epsilon_{\theta\theta} &= \frac{1}{r}\frac{\partial u_\theta}{\partial \theta}+\frac{u_r}{r}+\frac{1}{2}\left(\frac{1}{r}\frac{\partial w}{\partial \theta}\right)^2\,, \quad \epsilon_{rr} = \frac{\partial u_r}{\partial r}+\frac{1}{2} \left(\frac{\partial w}{\partial r}\right)^2\,,\\
    \epsilon_{r\theta}&=\epsilon_{\theta r}=\frac{1}{2}\frac{\partial u_\theta}{\partial r}-\frac{u_\theta}{2r}+\frac{1}{2r}\frac{\partial u_r}{\partial \theta}+\frac{1}{2}\frac{\partial w}{\partial r}\left(\frac{1}{r}\frac{\partial w}{\partial \theta}\right)\,.
\end{split}
\end{equation}
Here $w$ is the out-of-plane displacement of the panel, and $u_r$ and $u_\theta$ are the displacements in the radial and circumferential directions respectively. We assume a linear-elastic constitutive relationship between the stress, $\bm{\sigma}$, and the strain, $\bm{\epsilon}$, tensors:
 \begin{equation}
     \bm{\sigma} = \frac{E}{1-\nu^2}\left[ (1-\nu) \bm{\epsilon} + \nu \text{Tr}(\bm{\epsilon})\bm{I}\right]\,,
 \end{equation}
where $E$ is Young's modulus, $\nu$ is Poisson's ratio, and $\bm{I}$ represents the $2\times2$ identity matrix. Eqs.~\eqref{eq:FvK} above is solved perturbatively by choosing the following \emph{ansatz} for the displacement fields: $u_r(r,\theta) = \delta\,f(r)+\mathcal{O}\left(\delta^2\right)\,$, $u_\theta (r,\theta) = \delta\,r\theta\phi/\alpha_m+\mathcal{O}\left(\delta^2\right)$ and $w(r,\theta) = \delta\,r g(\theta)+\mathcal{O}\left(\delta^2\right)\,$, where $\delta$ is inserted as an artificial parameter in this perturbative series that has the sole purpose to keep track of the orders in the expansion.
Up to leading order, the \emph{ansatz} is such that the radial displacement is azimuthally symmetric, the circumferential deformation is uniform across the panel, and the out-of-plane deflection grows linearly along the radial direction. Furthermore, by considering this problem up to the leading order, stretching and bending are decoupled and these may be solved separately: Eq.~\eqref{eq:FvK1} yields a second order ODE for $f(r)\,$, Eq.~\eqref{eq:FvK2} is trivially satisfied, and Eq.~\eqref{eq:FvK3} yields a fourth order ODE for $g(\theta)\,$.

Under these conditions, looking at the stretching of the panel, Eq.~\eqref{eq:FvK1} leads to the following differential equation for the radial displacement, $f(r)$:
\begin{align}
f''(r) + \frac{1}{r}f'(r)-\frac{1}{r^2}f(r)-\frac{1-\nu}{r}\frac{\phi}{\alpha_m}=0\,. \label{eq:urEqn}
\end{align}
The above equation is solved subject to zero stress boundary conditions at the radial edges of the panel ($\sigma(R_i) = \sigma(R_o) = 0$) which yields
\begin{equation}
f(r)\!=\!\left[\frac{(1+\nu) R_i^2 R_o^2 \ln \left(\frac{R_o}{R_i}\right)}{r^2 \left(R_i^2-R_o^2\right)}+\frac{(1-\nu ) \left[R_i^2 \ln \left(\frac{r}{R_i}\right)-R_o^2 \ln \left(\frac{r}{R_o}\right)\right]}{ \left(R_i^2-R_o^2\right)}-1\right]\frac{r \phi}{2\alpha_m}\,.
\end{equation}
Therefore, the equivalent joint stiffness, $k_s\,$, is computed by equating the joint's rotational spring energy, $k_s \phi^2/2\,$,  to the linearised stretching energy cost of the panel:
\begin{equation}
k_s = \frac{h}{\phi^2}\int_0^{\alpha_m}\!\!\!\!\int_{R_i}^{R_o} \!\! \text{Tr}\left(\bm{\sigma} \cdot \bm{\epsilon}\right)r\,\mathrm{d}r\mathrm{d}\theta
= \frac{E h m}{8 \pi}\frac{ \left(R_i^2-R_o^2\right)^2-4 R_i^2 R_o^2 \left[\ln\left(\frac{R_o}{R_i}\right)\right]^2}{ (R_o^2-R_i^2)}\,.
\end{equation}

Next, we consider the moment balance in the panel. Eq.~\eqref{eq:FvK3} leads to the following differential equation for the function $g(\theta)$:
\begin{align}
g^{IV}(\theta ) + 2 g''(\theta )+g(\theta )=0\,. \label{eq:psiEqn}
\end{align}
This is solved subject to the panel edges remaining in plane ($g(0)=g(\alpha_m)=0$) and the imposed edge rotations ($g'(0)=-g'(\alpha_m)=\gamma/2$); its solution reads:
\begin{equation}
g(\theta)=\left(\frac{\theta  \sin (\alpha_m -\theta )+(\alpha_m -\theta ) \sin{\theta}}{\alpha_m +\sin{\alpha_m}}\right)\frac{\gamma}{2}\,.\label{eq:urExp}
\end{equation}
The equivalent hinge stiffness, $k_b\,$, is computed by equating the hinge's rotational spring energy, $k_b \gamma^2/2\,$, to the linearised bending energy of the panel:
\begin{equation}
k_b = \frac{D}{\gamma^2}\int_0^{\alpha_m}\!\!\!\!\int_{R_i}^{R_o}\!\!\left(\nabla^2w\right)^2r\; \mathrm{d}r\mathrm{d}\theta
=D\frac{1+\cos{(\pi/m)}}{\pi/m+ \sin{(\pi/m)}}\ln{\left(\frac{R_o}{R_i}\right)}\,,
\end{equation}
where $D=Eh^3/[12(1-\nu^2)]$ is the flexural rigidity of the plate.

We now have a complete mechanical description of the creased kirigami disk. In order to obtain the admissible equilibrium configurations, we minimize the total elastic energy $\mathcal W\,$, given in Eq.~\eqref{tot_E}, over the space of kinematically compatible deformations, i.e., subject to the continuity constraints given by Eq.~\eqref{kin_con}.

\section{The isometric case: creased model versus smooth analytical solution}
\label{sec:3}

In part I of this series \cite{sadik2021kiriI}, we concerned ourselves with the isometric solution of the e-cone\textemdash since the stretching energy contribution becomes negligible moving towards this limit\textemdash and analytically solved the continuum post-buckling problem in a geometrically nonlinear setting. Before investigating the influence of stretching deformations that are allowed using the present creased model, we first compare to the analytical solution by discarding the stretching contribution to the model, i.e., $\phi=0\,$. Thus we minimise the bending energy $\mathcal W_b\,$, as in Eq.~\eqref{tot_E}, with respect to the folding angles $\{\gamma^r_1,...,\gamma^r_{m-1},\gamma^l_1,...,\gamma^l_{m-1},\}$ subject to the kinematic constraints, Eq.~\eqref{kin_con}.

\begin{figure}
\centering
\def\svgwidth{.75\textwidth}
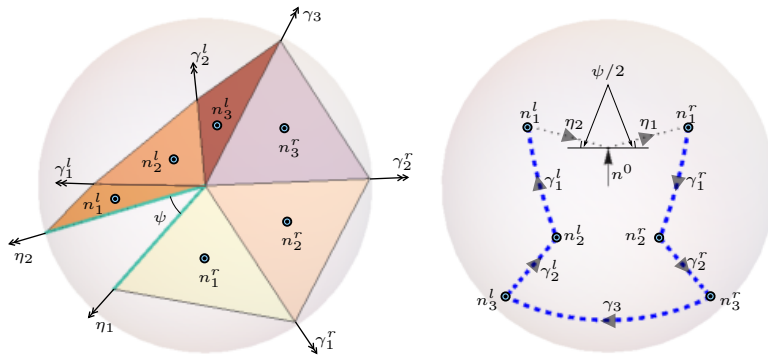
\caption{Gauss map representation of a creased e-cone for $m=3\,$. Each facet's unit normal (left) is represented by its embedding into the unit sphere $S^2$ (right). Spherical geodesics connect adjacent facets' normals on $S^2$ following the folding direction to form a piecewise smooth curve: each portion of the curve has a length equal to the folding angle it represents; and any two portions of the curve form a spherical angle that is equal to the solid angle made by the respective folding lines they represent (which in general is not equal to the panels' subtended angle $\alpha_m$ unless $\phi=0$\textemdash in the absence of any stretching).}
\label{fig:G_map}       
\end{figure}

In order to conveniently visualise the geometry of the deformed e-cones and compare the creased to the analytical solutions, we use the Gauss map representation \cite{farmer2005geometry, seffen2016fundamental, walker2018shape}\textemdash see Fig.~\ref{fig:G_map}. For any given surface, the Gauss map representation consists of the image of the Gauss-Weingarten map: a mapping which associates every point of the surface to points on the unit sphere $S^2$ through the normal field, i.e., $\mathbf{N}:\mathcal{S}\longrightarrow S^2\,$, where $\mathbf{N}$ is the normal field on a surface $\mathcal{S}\,$.
For the creased solutions, the outward unit normal of each rigid panel is represented as a point on the sphere; points representing two adjacent panels are connected by the spherical geodesic following the folding direction; subsequently, connecting all adjacent panels in this way produces the Gauss map representation as a piecewise smooth spherical curve on $S^2$ as seen on Fig.~\ref{fig:G_map}.
For the analytical conical solutions, the outward unit normal is the same along any given radial generator; which yields the Gauss map representation as a smooth curve on $S^2$ traced by the smoothly connected generators' normals as one sweeps the e-cone circumferentially.
A point of visual comparison may thus be given by super-imposing the Gauss map representations of the creased and analytical smooth solutions.

\begin{figure}
\centering
\def\svgwidth{0.8\textwidth}
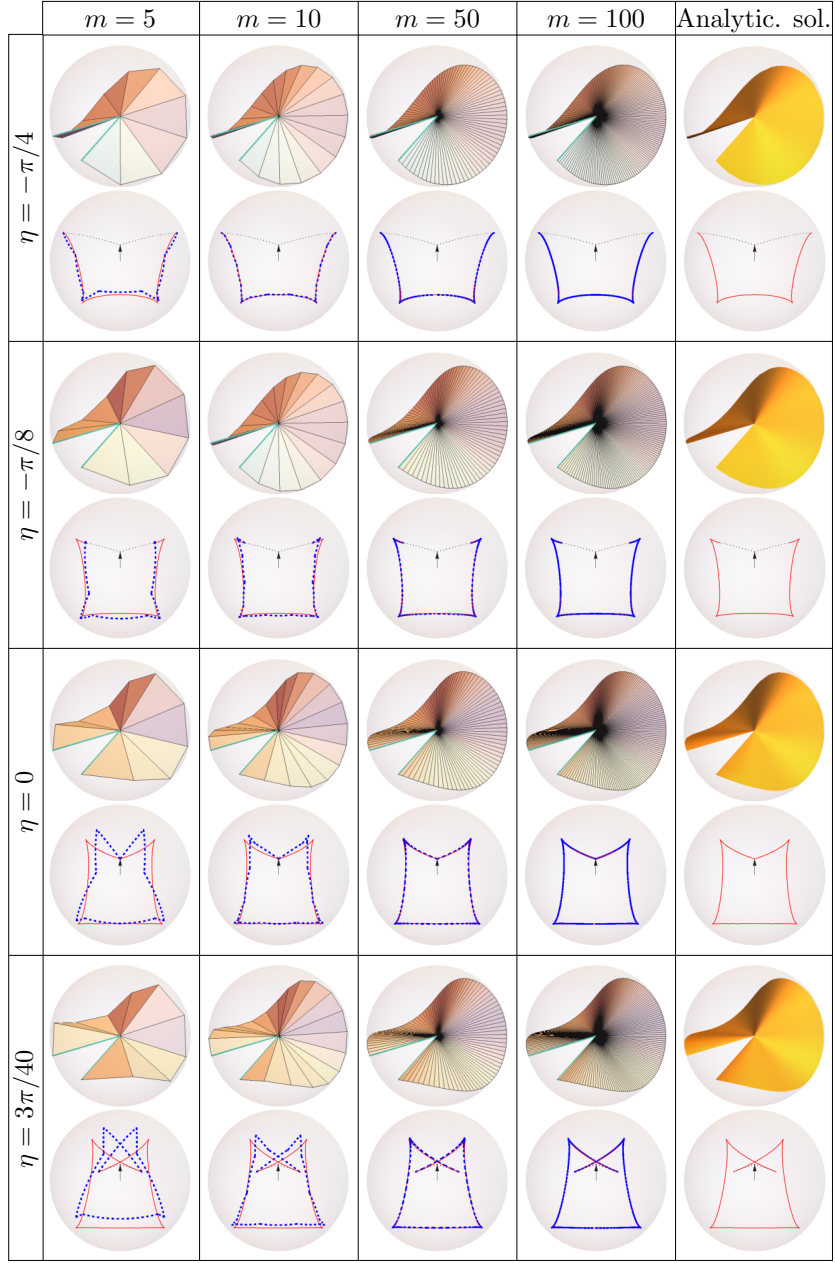
\caption{Gauss map representations of the creased solution for $m=5\,,\,10\,,\,50\,,\,\&\;100$ compared to the analytical results \cite{sadik2021kiriI} for an excess angle of $\psi=\pi/5$ and lips rotations of $\eta=-\pi/4\,,\,-\pi/8\,,\,0\,,\,\&\;3\pi/40\,$. The creased solution's Gauss maps quickly approach the corresponding analytical solution, even for a relatively small discretisation size, $m$.}
\label{fig:GM_comp}       
\end{figure}

Assuming mirror symmetric boundary conditions on the lips of the e-cone, i.e., $\eta=\eta_1=-\eta_2\,$, we solve for the symmetric configurations of the e-cone using both the creased and analytical approaches. In Fig.~\ref{fig:GM_comp}, for an excess angle of $\psi=\pi/5\,$, we show the e-cones obtained from the creased and analytical solutions, and their corresponding Gauss map representations for three different lip rotations $\eta$ ($-\pi/4\,$, $-\pi/8\,$, and $3\pi/40$). The creased solutions show a clear pattern of asymptotic convergence towards the analytical solutions with increasing discretisation size $m\,$.

\begin{figure}
\centering
\def\svgwidth{.75\textwidth}
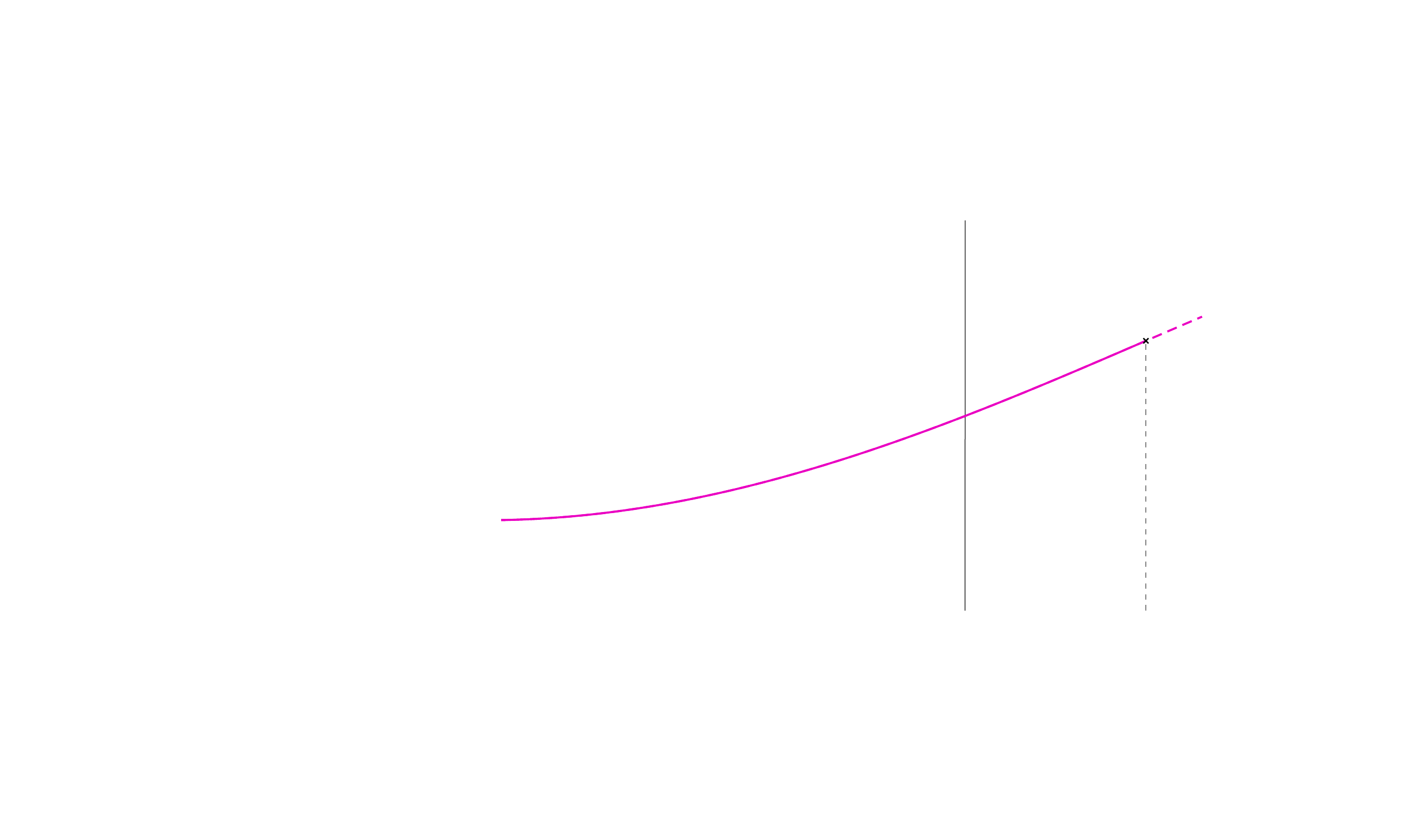
\caption{The onset of instability for the e-cone's symmetric creased configurations approaches that of the analytical smooth solution \cite{sadik2021kiriI} for an increasing disretisation size $m\,$. Here, the excess angle $\psi$ is fixed at $\pi/5\,$. In the inset, we see the onset of instability, $\eta_c\,$, i.e., the limit between the stable and unstable configurations: it is a point of inflection for the bending energy as a function of the lips rotation $\eta\,$.}
\label{fig:eta_c}       
\end{figure}

As previously discussed in Part I of this work \cite{sadik2021kiriI}, for a specific excess angle, $\psi\,$, and mirror symmetric boundary conditions, $\eta=\eta_1=-\eta_2\,$, there may be up to two symmetric stable states (corresponding to the two symmetric stable solution orbit branches) and multiple symmetric unstable states (corresponding to the unstable winding continuation of the aforementioned orbit branches).
Tracking the symmetric configurations of the e-cone for the lips rotation $\eta=\eta_1=\eta_2$, an inflection point of the bending energy is found at $\eta=\eta_c\,$, marking the onset of instability separating the stable from the unstable configurations along the orbit\textemdash see inset of Fig.~\ref{fig:eta_c}. Beyond this point, i.e., for a lips' rotation $|\eta|>|\eta_c|\,$, the e-cone is in an unstable configuration and may experience a snap-through transition to the corresponding stable configuration for the same lips' rotation $\eta\,$. Plotting the value of the critical lips rotation $\eta_c$ of the creased solutions, against the discretisation size $m\,$, it may be seen that it asymptotically approaches, for increasing $m\,$, the critical lips rotation $\eta_c$ for the analytical solution\textemdash see Fig.~\ref{fig:eta_c}.

\begin{figure}[t]
\centering
\def\svgwidth{\textwidth}
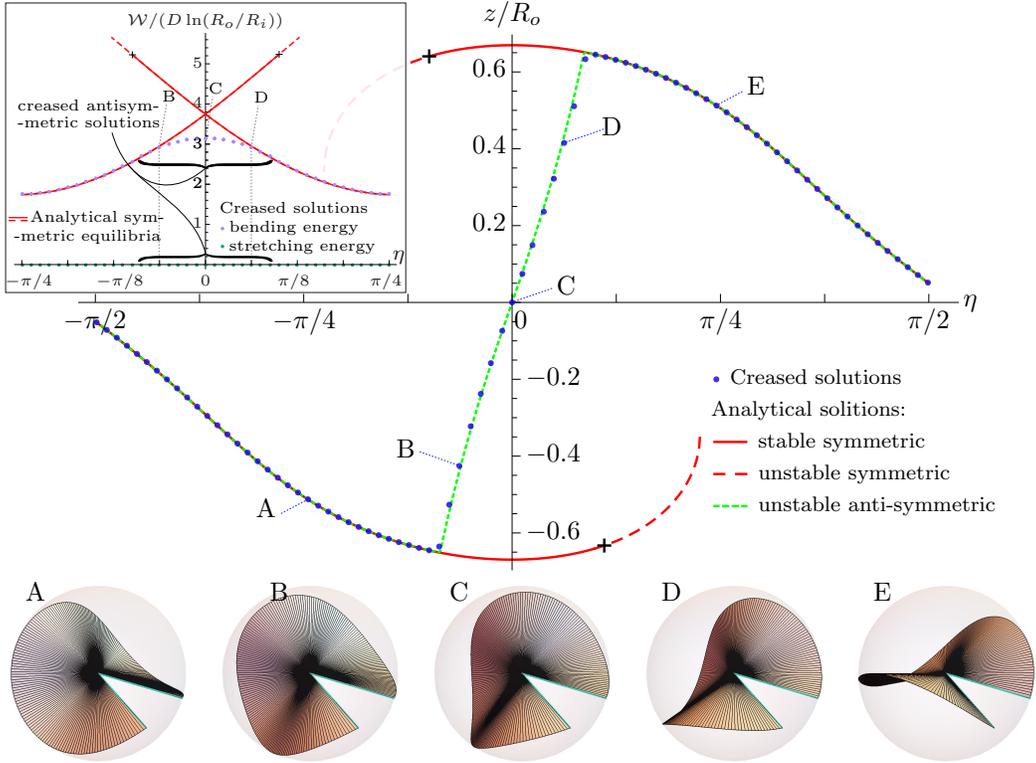
\caption{Symmetric solution orbits and anti-symmetric connecting path for both the  creased (m=100) and analytical \cite{sadik2021kiriI} solutions. The solutions are shown in the $\{z/R_o,\eta\}$ phase space for $h/R_o=10^{-3}\,$, with $z$ being the deflection of the antipodal point to the slit. The inset shows the energy landscape for both the creased and analytical solutions. The $\bm{+}$ indicates the limit of stability for the symmetric solution.}
\label{fig:z_Nn_Eq}       
\end{figure}

Fig.~\ref{fig:z_Nn_Eq} shows the symmetric stable solution orbits, which are connected by an unstable anti-symmetric path. We represent these solutions in the $\{\eta,z/R_o\}$ phase space, where $z$ is the deflection of the antipodal point to the slit. The creased solution (for $m=100$) closely matches the analytical solution along the two stable symmetric paths, and the unstable anti-symmetric path connecting them. Stretching in the creased model is included to provide a smoothing effect, which facilitates the numerical scheme to find the global minimum solutions. However, when the total strain energy is minimised the stretching energy is found to be vanishingly small. The inset of Fig.~\ref{fig:z_Nn_Eq} shows the energy landscape for the e-cone, the creased solution shows the lower-energy anti-symmetric path between the two symmetric stable orbits.

\section{Beyond isometry: e-cones of finite thickness}
\label{sec:4}

We now include the stretching energy contribution in the creased model of the e-cone\textemdash that is, we minimise the total energy $\mathcal W$ over the space of kinematically compatible deformations\textemdash and consider its influence near the threshold of instability. Beyond providing a computationally efficient method to solve the isometric problem (as shown indeed in \S\ref{sec:2}\textemdash the problem is essentially reduced to a constrained algebraic minimisation procedure), our creased model allows us to further study the mechanics of finite thickness plates, where the stretching energy contribution can no longer neglected.

\begin{figure}
\centering
\def\svgwidth{\textwidth}
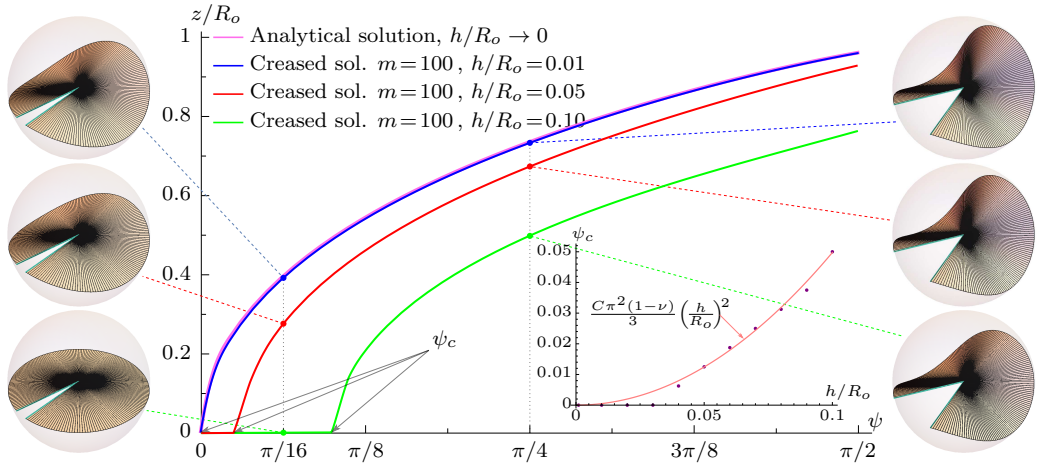
\caption{Deflection of the antipodal point to the slit for increasing values of the excess angle, $\psi\,$, plotted for different thicknesses of the creased model with $m=100\,$. The near-threshold behaviour of finite thickness e-cones at the critical excess angle $\psi_c$ can be clearly observed: unlike the isometric case where $\psi_c=0\,$, out-of-plane buckling is delayed for increasing thicknesses. In the inset, the critical excess angle, $\psi_c\,$, is plotted against the dimensionless thickness\textemdash data fitting identifies the quadratic scaling $\psi_c \propto (h/R_o)^2\,$.}
\label{fig:z_bif}       
\end{figure}

Fig.~\ref{fig:z_bif} shows the deflection of the antipodal point to the slit in the creased e-cone model with $m=100$ for increasing values of the excess angle, $\psi$ (at a fixed $\eta=0$). In the isometric case, $h/R_o \rightarrow 0\,$, out-of-plane buckling occurs for any excess angle $\psi>0\,$. However, as the thickness of the plate increases, the critical excess angle, $\psi_c\,$, at which out-of-plane buckling occurs, also increases. Indeed, as shown by the side snapshots of Fig.~\ref{fig:z_bif}, below the threshold of the critical excess angle, the Gaussian charge due to the imposed excess angle is fully consumed by in-plane stretching deformations before eventually buckling out-of-plane at the critical threshold $\psi=\psi_c\,$. The inset shows the dependence of the critical excess angle, $\psi_c\,$, on the dimensionless thickness. Fitting the results for the critical excess angle against the thickness reveals $\psi_c\approx C\pi^2(1-\nu)(h/R_o)^2/3$ where $C\approx 2.2$ is a geometric buckling coefficient\textemdash cf. to the critical buckling strain of a rectangular plate, where a linear stability analysis of the F{\"o}ppl-von K{\'a}rm{\'a}n equations yields that $C=1$ \cite{audoly2010elasticity}.

\begin{figure}[t]
\centering
\def\svgwidth{.75\textwidth}
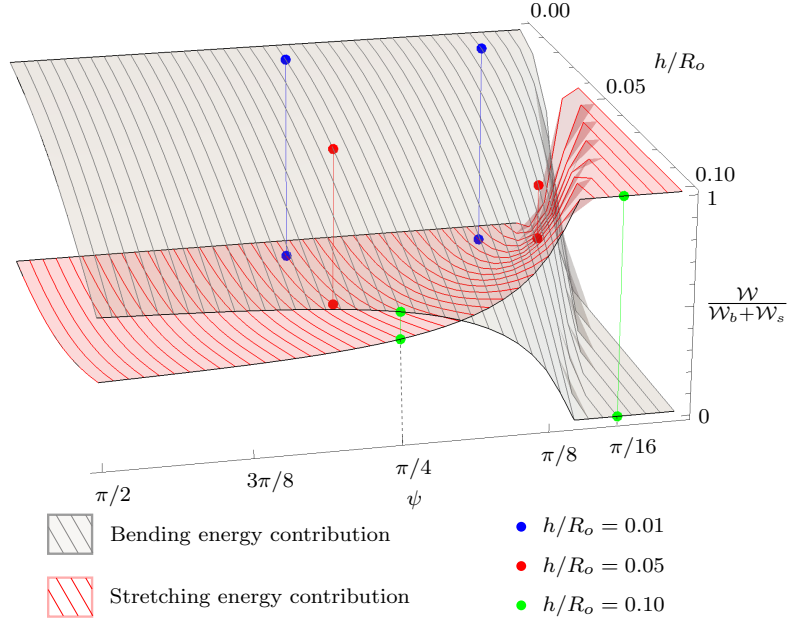
\caption{Strain energy equipartition showing bending and stretching contributions as a function of the excess angle, $\psi\,$, and the plate's dimensionless thickness, $h/R_o\,$, for a creased e-cone ($m=100$) with a fixed lip rotation $\eta=0\,$. Vertical lines connect the corresponding stretching and bending contributions for equilibrium states with $\psi=\pi/4,\, \pi/16$ and $h/R_o=0.01,\, 0.05,\,$ and $0.10$.}
\label{fig:E_part}       
\end{figure}

The equipartition of strain energy into bending and stretching contributions for increasing values of the excess angle, $\psi\,$, and the dimensionless thickness, $h/R_o\,$, is shown in Fig.~\ref{fig:E_part}. For the isometric case, $h/R_o \rightarrow 0\,$, the stretching contribution is nil, and the behaviour of the e-cone is solely governed by its bending energy\textemdash corresponding to out-plane-buckling immediately occurring when the slit is opened, i.e., $\psi_c=0$ (see Fig.~\ref{fig:z_bif}). As the thickness of the plate increases, the stretching contribution becomes increasingly more important to the point that it even overtakes the bending contribution. In fact, a region where the bending contribution becomes nil emerges for large thicknesses and small excess angles. This corresponds to the region below the threshold of the non-zero critical excess angle, $\psi<\psi_c\,$, seen in Fig.~\ref{fig:z_bif}.

\begin{figure}[!ht]
\centering
\def\svgwidth{.75\textwidth}
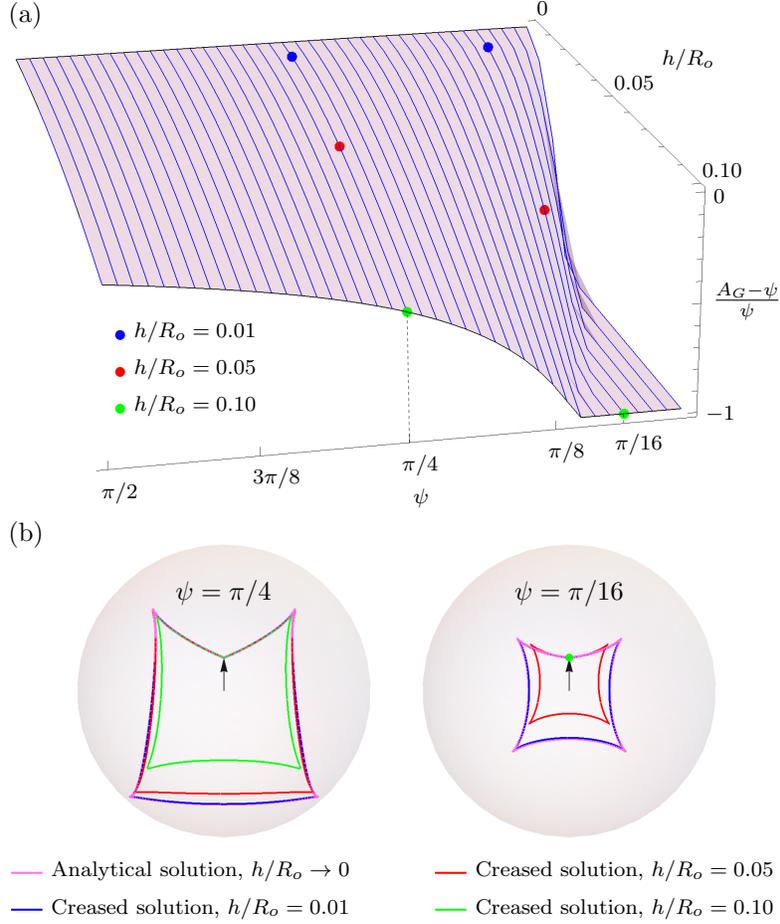
\caption{(a) Gaussian charge deficit expressed as the relative difference between available Gaussian charge, $A_G\,$, and the excess angle, $\psi\,$, as a function of the dimensionless thickness, $h/R_o\,$, and the excess angle, $\psi\,$  for a creased e-cone ($m=100$) with fixed lip rotation $\eta=0\,$. (b) Gauss map snapshots corresponding to points on (a) and Fig.~\ref{fig:E_part} for selected values of $\psi$ and increasing plate dimensionless thicknesses $h/R_o\,$. When the Gaussian charge is completely consumed by in-plane stretching, e.g., when $\psi=\pi/16$ and $h/R_o=0.1\,$, the Gauss map representation is reduced to a point.}
\label{fig:A_KG}       
\end{figure}

\begin{figure}
\centering
\def\svgwidth{.75\textwidth}
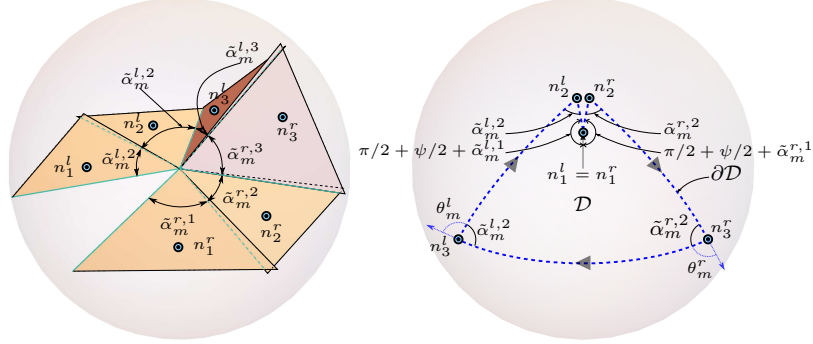
\caption{Gauss map representation of a creased ($m=3$) e-cone with $\psi=\pi/5$ and $\eta=0\,$ undergoing a combination of stretching and bending deformations. Unlike the Gauss map representation in Fig.~\ref{fig:G_map}, where the the e-cone is only subject to bending, here stretching forces the panels to overlap at the creases, effectively reducing the subtended angles of the panels and with it the area of the spherical surface, $\mathcal D$, enclosed by the path $\partial \mathcal D$ described by the surface normals.
}
\label{fig:G_map2}       
\end{figure}

Fig.~\ref{fig:A_KG} shows the relative difference between the effective Gaussian charge concentrated at the apex of the e-cone (the area, $A_G\,$, enclosed by the Gauss map \textemdash see Fig.~\ref{fig:G_map2}) and the excess angle $\psi\,$. Subjecting a flat disk to an excess angle introduces a Gaussian charge at the apex, which causes deformation of the disk. In the isometric limit, this charge fully translates into bending to form a stretch-free e-cone. In other words, the excess angle fully manifests itself as the effective Gaussian charge at the apex of the e-cone. In this case the total area of the Gauss map, $A_G$, is identical to the excess angle, $\psi\,$, as seen in Fig.~\ref{fig:A_KG} for $h/R_o \rightarrow 0\,$.
However, for finite thickness plates, a combination of bending and stretching is experienced, as seen in Fig.~\ref{fig:E_part}. Before reaching the initial instability threshold, the energy landscape is pure in-plane stretching. Therefore, for finite thickness plates, part, or all, of the Gaussian charge is consumed by in-plane stretching such that there exists a mismatch between the excess angle, $\psi\,$, and the available Gaussian charge, $A_G\,$\textemdash see Fig.~\ref{fig:A_KG}. Indeed, this mismatch turns into in-plane stretching, while the remaining available Gaussian charge turns into out-of-plane bending deformation. This explains the striking similarity between the relative Gaussian charge mismatch shown in Fig.~\ref{fig:A_KG} and the energy equipartition profile shown in Fig.~\ref{fig:E_part}.
Further, the Gauss map snapshots in Fig.~\ref{fig:A_KG} show how, for the same excess angle, $\psi\,$, the available Gaussian charge, $A_G\,$ (or equivalently the area enclosed by the Gauss map), is reduced as the thickness is increased.

In order to quantify the available Gaussian charge discussed above, we must compute the actual area, $A_G\,$, of the surface patch $\mathcal D$ with boundary $\partial \mathcal D$ given by the Gauss map representation of the e-cone\textemdash see Fig.~\ref{fig:G_map2}.
Let us first make a few important observations about the Gauss map representation of Fig.~\ref{fig:G_map2} in contrast to that of Fig.~\ref{fig:G_map}. Figure~\ref{fig:G_map} concerns an isometric (and hence stretch-free) e-cone, while Fig.~\ref{fig:G_map2} concerns a stretched e-cone. In both figures, the lengths of the geodesic edges composing the path are exactly equal to the folding angle of the corresponding crease. However, the external turning angle of the geodesics composing the path are different. In Fig.~\ref{fig:G_map}, since there is no stretching, the external turning angle of the geodesics composing the path are given by the panels' subtended angle $\alpha_m=\pi/m\,$. In contrast, for the stretched e-cone of Fig.~\ref{fig:G_map2}, the presence of compression forces the panels to overlap at the creases effectively changing their subtended angle to $\tilde\alpha^i_m$\textemdash the superscript (r - right, l - left) designates the side where the panel is located\textemdash depending on both the folding angle at the crease, $\gamma_i$, and the overlap angle, $\phi\,$. Consequently, the turning angle is different for each panel and given by the corresponding modified subtended angle $\tilde\alpha^i_m\,$.

We now apply the Gauss-Bonnet theorem to the surface patch $\mathcal D$ on the unit sphere in order to compute its surface area, $A_G\,$, which reads as follows:
\begin{equation}\label{G-B}
\int_{\mathcal D\subset S^2} K\,dA + \int_{\partial \mathcal D}k_g\,ds + \theta^1_m + \sum_{i=2}^m (\theta^{r,i}_m+\theta^{l,i}_m) = 2 \pi \chi(\mathcal D)\,,
\end{equation}
where $K$ is the Gaussian curvature of $\mathcal D\,$, which is equal to $1$ since $\mathcal D\subset S^2\,$; $k_g$ is the geodesic curvature of $\partial \mathcal D\,$, which is identically equal to zero since the spherical boundary $\partial \mathcal D$ traces a path composed of great circles on $S^2\,$; $\chi(\mathcal D)$ is the Euler characteristic of $\mathcal D\,$, which is equal to 1 since $\mathcal D$ is a simply connected patch; and $\theta^1_m=-\alpha^{r,1}_m-\alpha^{l,1}_m+\psi\,$ is the external turning angle associated to the panels by slit\textemdash see Fig.~\ref{fig:G_map2}. For $i\in\{2,...,m\}\,$, $\theta^i_m$ is the external turning angle associated with the $i^{th}$ panel, which depending on the direction of the fold, reads either as $\theta^i_m=\pi-\tilde\alpha^i_m$ or $\theta^i_m = -\tilde\alpha^i_m\,$, see Fig.~\ref{fig:G_map2}. It follows that Eq.~\ref{G-B} transforms to yield the area $A_G$
\begin{equation}\label{eq:AG_f}
A_G\equiv \psi - 2 \pi +\sum_{i=1}^m (\alpha^{r,i}_m+\alpha^{l,i}_m) \,.
\end{equation}
Note that in the isometric case, since $\alpha^{r,i}_m=\alpha^{l,i}_m=\alpha_m=\pi/m\,$, we have $\sum_{i=1}^m (\alpha^{r,i}_m + \alpha^{l,i}_m) = 2\pi$ and Eq.~\eqref{eq:AG_f} reduces to $A_G\equiv\psi\,$. That is, the Gaussian charge is exactly equal to the excess angle.
However, for stretched e-cones, we have $\sum_{i=1}^m (\alpha^{r,i}_m+\alpha^{l,i}_m)<2\pi$ and hence the Gaussian charge mismatch shown in Fig.~\ref{fig:A_KG}.

\section{Concluding remarks}

In this work, we develop a model which moves beyond the isometric limit to provide a more complete picture of local kirigami mechanics through its fundamental building block\textemdash the e-cone (as shown in Fig.~\ref{fig:Kiri2Econe}). Our new model fully captures the entire range of far-from-threshold post-buckling behaviour as well as enabling investigations of near-threshold instabilities, which must account for both the effects of both bending and stretching. The model discretises the e-cone using a series of radial creases connecting kinematically flat rigid panels. The creases allow both relative rotation and separation between panels, capturing bending and stretching deformations, respectively. To identify admissible equilibrium configurations, these deformations are penalised by elastic springs with stiffnesses derived from continuum deformations of an elastic plate with the crease deformations imposed as boundary conditions. The fidelity of the model is improved simply by increasing the number of creases, formally approaching the continuum in the limit of large numbers of creases.

For finite-thickness sheets, our model shows that for increasing thicknesses the Gaussian charge, as a result of an imposed excess angle, is partially consumed by in-plane stretching, thus reducing the magnitude of out-of-plane deformations. For sufficiently thick sheets, this charge can be completely absorbed into pure in-plane stretching, i.e., no out-of-plane deformations occur. This phenomenon arises in connection with what is expected from the continuum theory, whereby one of the F{\"o}ppl-von K{\'a}rm{\'a}n equations relates the curvature charge to the Airy stress potential as a source term~\cite{moshe2019kirigami,seung1988defects}. Moreover, thickness also has the effect of increasing the critical excess angle for the onset of buckling, from zero in the isometric limit and increasing with the thickness squared. Understanding this effect is essential for engineering applications of kirigami structures, which depends on identifying the conditions required to transition from in-plane to out-of-plane deformations and the corresponding switch from low to high stretchability. As such, this local approach provides the required understanding of the building block, leading to better control of effective mechanical properties in kirigami structures, as well as other desirable properties\textemdash e.g., shape-changing and multistability. 

By building on Part I of this series, this work further underpins the phenomenology of kirigami structures with a proper theoretical approach to geometry and mechanics. It will enable better designs and reduce computational costs. Our analysis approach offers the computational efficiencies of reduced order bending models while also capturing stretching and formally approaching continuum behaviour in the limit. By identifying the underlying kinematic behaviour, this approach can also be applied to other large-deformation shell mechanics problems and offers the potential to improve our understanding of the behaviour of thin-walled structures more generally.

\section{Acknowledgements} 
SS and MAD would like to thank the Velux Foundations for support under the Villum Experiment program (Project No. 00023059).

\bibliography{ref}

\end{document}